\documentclass[12pt]{article}
\usepackage{epsfig}
\newcommand{\beq}{\begin{equation}}
\newcommand{\eeq}{\end{equation}}
\newcommand{\beqa}{\begin{eqnarray}}
\newcommand{\eeqa}{\end{eqnarray}}
\newcommand{\beqar}{\begin{eqnarray*}}
\newcommand{\eeqar}{\end{eqnarray*}}

\begin{document}
\addtolength{\baselineskip}{1.5mm}

\thispagestyle{empty}


\vspace{32pt}

\begin{center}
\textbf{\Large\bf Quantum Description for the Decay of NSNS Brane-Antibrane Systems}

\vspace{48pt}

Hongsu Kim\footnote{hongsu@astro.snu.ac.kr}

\vspace{12pt}

\textit{Astronomy Program, SEES, Seoul National University \\
Seoul, 151-742, KOREA}

\end{center}
\vspace{48pt}

\begin{abstract}

The stringy description for the instabilities in the $RR$ charged $D_{p}-\bar{D}_{p}$ pairs is 
now well understood in terms of the open string tachyon condensation. The quantum interpretation 
presumably via the stringy description for the instabilities in the $NSNS$-charged $F1-\bar{F1}$ 
and $NS5-\bar{NS5}$ pairs in IIA/IIB theories, however, has not been fully established yet. 
This would be partly because 
of the absence (for the $F1-\bar{F1}$ case) or our relatively poor understanding 
(for the $NS5-\bar{NS5}$ case) of their worldvolume (gauge theory) dynamics. In the present work, 
using the well-known quantum description for instabilities in the $RR$-charged $D_{p}-\bar{D}_{p}$ 
systems and in the M-theory brane-antibrane systems and invoking appropriate string dualities, 
the stringy nature of the instabilities in the $NSNS$-charged $F1-\bar{F1}$ and $NS5-\bar{NS5}$ 
systems both at strong and at weak couplings has been uncovered. 
For the annihilations to string vacua, the quantum, stringy interpretations are simple extensions 
of Sen's conjecture for those in $RR$-charged brane-antibrane systems.

\end{abstract}

\hfill{PACS numbers : 11.25.Sq, 04.65.+e, 11.10.Kk}

\setcounter{footnote}{0}

\newpage

\section{Introduction}

The stringy description for the instabilities in the $RR$ charged $D_{p}-\bar{D}_{p}$ 
pairs \cite{non-BPS} is 
now well understood in terms of the open string tachyon condensation. Consider a system consisting of a certain number $N$ of coincident $D_{P}$-branes separated by some distance from a
system of $N$ coincident $\bar{D}_{p}$-branes, for simplicity, in flat $R^{10}$. This system differs from the BPS system of $2N$
$D_{p}$-branes by the orientation reversal on the antibranes. In this system, the branes and the antibranes each break a different
half of the original supersymmetry and the whole configuration is non-supersymmetric or non-BPS and hence is unstable. As a result,
there is a combined gravitational and (RR) gauge attractive force between the branes and the antibranes at some large but finite
separation leading to the semi-classical instability. At the separation of order the string scale $\sim \sqrt{\alpha'}=l_{s}$,
in particular, the open string connecting a $D_{p}$-brane to a $\bar{D}_{p}$-brane becomes tachyonic. What then would be the
eventual fate or endpoint of this unstable $D_{p}-\bar{D}_{p}$-system ? According to Sen \cite{sen}, the endpoint could be the
supersymmetric vacuum via the open string tachyon condensation. Later on in section 2.2, we shall be
more specific on this. The quantum interpretation, again presumably via the stringy description for 
the instabilities in the $NSNS$-charged $F1-\bar{F1}$ and $NS5-\bar{NS5}$ 
pairs in IIA/IIB theories, however, has not been fully established yet \cite{kim}. 
This would be partly due 
to the absence (for the $F1-\bar{F1}$ case) or our relatively poor understanding 
(for the $NS5-\bar{NS5}$ case) of their worldvolume (gauge theory) dynamics. (Later on in the
concluding remarks, we shall summarize particularly some of the collected wisdom on the nature 
of $NS5$-brane worldvolume dynamics uncovered thus far.) 
And certainly, this nature is reflected in the intersection rules saying that
fundamental strings do not end on themselves nor on $NS5$-branes, i.e., neither $(0|F1,F1)$ nor
$(0|F1,NS5)$ exists. As a result, unlike the $RR$-charged $D_{p}$-brane case in which the fundamental
strings ending on $D_{p}$ or $\bar{D}_{p}$ essentially provides the worldvolume dynamics of the branes
and generates the instabilties in the $D_{p}-\bar{D}_{P}$ systems, in this $NSNS$-charged $F1-\bar{F1}$
and $NS5-\bar{NS5}$ systems, the quantum, elementary entity both to provide the worldvolume dynamics
and to generate the instabilities in the brane-antibrane systems is missing \cite{kim}. 
This is certainly an 
embarrassing state of affair if we realize the fact that these $NSNS$-charged brane-antibrane pairs
are just $U$-duals of $RR$-charged ones for which the quantum, stringy description for the
instabilities is well-established in terms of Sen's conjecture of open string tachyon 
condensation \cite{sen}. In the present work, therefore,
using the well-known quantum description for instabilities in the $RR$-charged $D_{p}-\bar{D}_{p}$ 
systems and in the M-theory brane-antibrane systems and invoking appropriate string dualities, 
the stringy nature of the instabilities in the $NSNS$-charged $F1-\bar{F1}$ and $NS5-\bar{NS5}$ 
systems both at strong and at weak couplings has been uncovered. 
And particularly for the annihilations to string vacua, the quantum, stringy 
interpretations are simple extensions of Sen's conjecture given for those in $RR$-charged 
brane-antibrane systems.

\section{Instabilities in IIB theory ($F1-\bar{F1}$, $NS5-\bar{NS5}$) systems}\label{ }

\subsection{Supergravity description of the instability}

In this section, we would like to demonstrate in an explicit manner that via the $S$-duality, one 
can actually obtain the supergravity solutions representing $F1-\bar{F1}$ and $NS5-\bar{NS5}$ systems
from those representing $D1-\bar{D1}$ and $D5-\bar{D5}$ systems rspectively. Thus to this end, we begin
with the type IIB supergravity action in string frame
\beqa
S_{IIB} &=& {1\over 2\kappa^2}\int d^{10}x\sqrt{g} \left\{e^{-2\phi}\left[R + 4(\nabla \phi)^2
- {1\over 12}H^2_{[3]}\right] - {1\over 2}(\partial \chi)^2\right. \\
&-& \left. {1\over 12}F^2_{[3]} - {1\over 240}F^2_{[5]}\right\} +
{1\over 4\kappa^2}\int A_{[4]}\wedge dA_{[2]}\wedge H_{[3]} \nonumber
\eeqa
where $\kappa^2 = 8\pi G$, $H_{[3]}=dB_{[2]}$ is the field strength of the $NSNS$ tensor field
$B_{[2]}$ and $F_{[3]}=dA_{[2]}-\chi H_{[3]}$, $F_{[5]}=dA_{[4]}+A_{[2]}\wedge H_{[3]}$ are the
$RR$ field strengths. And when one writes the type IIB supergravity action as above, it is implicit
that one never asks that the self-(Hodge) duality condition on the $5$-form $RR$ field strength
$F_{[5]}$ follows from the variation of this action but instead is assumed to be
imposed afterwards by hand, $*F_{[5]} = F_{[5]}$. 
Now note that the IIB theory equations of motion that result by extremizing
the IIB theory action given above are invariant under a $SL(2,R)$ symmetry (and it is broken to
$SL(2,Z)$ in the full type IIB string theory) under which the fields transform as follows.
Namely define 
\beq
\lambda = \chi + ie^{-\phi}, ~~~H=\pmatrix{ B_{[2]} \cr A_{[2]} \cr  }
\eeq
where $\chi$ and $\phi$ are the $RR$ scalar (axion) and the $NSNS$ scalar (dilaton) respectively
appeared in the IIB theory action given above. Then, under a $SL(2,R)$ transformation represented
by the matrix
\beq
U = \pmatrix { a & b \cr 
               c & d \cr 
     } \in SL(2,R) ~~~{\rm with} ~~~ad-bc = 1,
\eeq
the scalars and the 2-form potentials transform according to
\beq
\lambda \rightarrow {a\lambda + b \over c\lambda + d},
~~~H \rightarrow UH = \pmatrix { a & b \cr  c & d \cr  }H
\eeq
with the other fields remaining invariant. Now, consider a particular case of this $SL(2,R)$
transformation in which one sets the $RR$ scalar (axion) to zero, $\chi = 0$. This particular
$SL(2,R)$ transformation amounts to choosing ($a=0$, $b=1$, $c=-1$, $d=0$) under which the
fields transform as
\beq
\lambda' = -{1\over \lambda} ~({\rm or} ~e^{-\phi'} = e^{\phi}),
~~~\pmatrix{ B'_{[2]} \cr A'_{[2]} \cr } = \pmatrix { A_{[2]} \cr -B_{[2]} \cr}, ~~~g'_{\mu\nu} = e^{-\phi}g_{\mu\nu}
\eeq
(in the string frame) and is referred to as {\it S-duality}. \\
With this preparation, we first start with the supergravity solution representing the 
$D1-\bar{D1}$ pair of type IIB theory \cite{youm}.
\beqa
ds^2_{10} &=& H^{-1/2}[-dt^2 + dx^2_{1}] +
H^{1/2}[\sum^{6}_{m=2} dx^2_{m} + (\Delta+a^2\sin^2 \theta)\left({dr^2\over
\Delta}+d\theta^2\right) + \Delta \sin^2 \theta d\phi^2], \nonumber \\ 
e^{2\phi} &=& H, \\ 
A_{[2]} &=& \left[{2ma\cos \theta \over \Sigma}\right]dx_{1}\wedge dt, ~~~F_{[3]}=dA_{[2]} 
\nonumber 
\eeqa 
where the ``modified'' harmonic function is given by $H(r) = \Sigma /({\Delta + a^2\sin^2 \theta})$
and $\Sigma = r^2-a^2\cos^2 \theta$, $\Delta = r^2 - 2mr - a^2$ with $m$ being the ADM mass of each
$D$-brane. Then the ADM mass of the whole $D1-\bar{D1}$ system is $M_{ADM}=2m$ which should be
obvious as it would be the sum of ADM mass of each brane when they are well separated. The parameter
$a$ can be thought of as representing the proper distance between 
the brane and the antibrane \cite{kim}.
Now, applying the $S$-duality transformation laws given above to this $D1-\bar{D1}$ solution, one
gets the following supergravity solution
\beqa
ds^2_{10} &=& H^{-1}[-dt^2 + dx^2_{1}] + \sum^{6}_{m=2}dx^2_{m} +
(\Delta+a^2\sin^2 \theta)\left({dr^2\over
\Delta}+d\theta^2\right) + \Delta \sin^2 \theta d\phi^2, \nonumber \\ 
e^{2\phi} &=& H^{-1}, \\ 
B_{[2]} &=& \left[{2ma\cos \theta \over \Sigma}\right]dx_{1}\wedge dt,
~~~H_{[3]} = dB_{[2]} \nonumber 
\eeqa 
which is indeed the $F1-\bar{F1}$ solution of IIA/IIB theory \cite{youm}. \\
Next, we start with the supergravity solution representing the $D5-\bar{D5}$ pair of 
IIB theory \cite{youm}.
\beqa
ds^2_{10} &=& H^{-1/2}[-dt^2 + \sum^{5}_{i=1}dx^2_{i}] +
H^{1/2}[dx^2_{6} + (\Delta+a^2\sin^2 \theta)\left({dr^2\over
\Delta}+d\theta^2\right) + \Delta \sin^2 \theta d\phi^2], \nonumber \\ 
e^{2\phi} &=& H^{-1}, \\ 
A^{m}_{[2]} &=& \left[{2mra\sin^2 \theta \over \Delta + a^2 \sin^2 \theta}\right]dx_{6}\wedge d\phi, 
~~~F^{m}_{[3]} = dA^{m}_{[2]}. \nonumber 
\eeqa
Similarly, by applying the $S$-duality transformation law to this $D5-\bar{D5}$ solution, one gets
the following supergravity solution
\beqa
ds^2_{10} &=& -dt^2 + \sum^{5}_{i=1}dx^2_{i} +
H[dx^2_{6} + (\Delta+a^2\sin^2 \theta)\left({dr^2\over
\Delta}+d\theta^2\right) + \Delta \sin^2 \theta d\phi^2], \nonumber \\ 
e^{2\phi} &=& H, \\ 
B^{m}_{[2]} &=& \left[{2mra\sin^2 \theta \over \Delta + a^2 \sin^2 \theta}\right]dx_{6}\wedge d\phi, 
~~~H^{m}_{[3]} = dB^{m}_{[2]} \nonumber 
\eeqa
which indeed can be identified with the $NS5-\bar{NS5}$ solution of IIA/IIB theory \cite{youm}.
That these IIB theory $D1-\bar{D1}$, $D5-\bar{D5}$ pairs and their $S$-dual $F1-\bar{F1}$,
$NS5-\bar{NS5}$ pairs represented by the supergravity solutions given above indeed exhibit
semi-classical instabilities in terms of the appearance of the {\it conical singularities}
can be found in detail in our earlier work \cite{kim} in which $D6-\bar{D6}$ system has been 
taken for explicit demonstration. Indeed, the nature of semi-classical instabilities owned by these
$RR$/$NSNS$-charged brane-antibrane systems in terms of the conical singularities is very
reminiscent of essentially the same conical singularity structure in the Bonnor's magnetic 
dipole solution \cite{bonnor} in Einstein-Maxwell theory. We now attempt to describe briefly the
nature of these semi-classical instabilities. consider the symmetry axis $\theta=0, \pi$ connecting
the brane and the antibrane. Then one can see that the conical singularities arise both along the
semi-infinite axes extending from the (anti)brane to infinity and along the line segment between
the brane and the antibrane. One, however, can immediately realise that both of the two conical
singularities cannot be eliminated at the same time. Then the usual option one takes is to remove
the conical singularity along the line segment between the two at the expense of the appearance of
the conical angle deficits along the semi-infinite axes. This, in turn, implies the presence of
the {\it cosmic strings} providing tension that pulls the brane and the antibrane apart against
the collapse due to the combined gravitational and gauge attractions. Nevertheless, the 
semi-classical instabilities owned by these brane-antibrane systems still manifest themselves
since it is not hard to realise that the cosmic strings can only suspend the brane-antibrane
systems in an {\it unstable} equilibrium configuration. Namely, it can be demonstrated that this
unstable equilibrium is indeed vulnerable since if one brings the brane and the antibrane close
to each other, they always collide and merge completely. To see this, note first that the parameter
$a$ appearing in the supergravity solutions above can be regarded as representing the proper
separation between the brane and the antibrane \cite{kim}. \\
Now consider the $D1-\bar{D1}$ solution given in eq.(6). In the limit $a\rightarrow 0$, it becomes
\beqa
ds^2_{10} &=& \left(1 - {2m\over r}\right)^{1/2}[-dt^2 + dx^2_{1}] \nonumber \\
&+& \left(1 - {2m\over r}\right)^{-1/2}[\sum^{6}_{m=2} dx^2_{m} + dr^2
+ r^2\left(1 - {2m\over r}\right)(d\theta^2 + \sin^2 \theta d\phi^2)], \nonumber \\ 
e^{2\phi} &=& \left(1 - {2m\over r}\right)^{-1}, ~~~A_{[2]} = 0  
\eeqa 
where we used $\Sigma \rightarrow r^2$, $\Delta \rightarrow r^2(1-2m/r)$, and hence 
$H \rightarrow (1-2m/r)^{-1}$ as $a\rightarrow 0$. In this limit, the opposite $RR$ charges
carried by $D1$ and $\bar{D1}$ annihilated each other since $A_{[2]}=0$ and the solution
now has the topology of $R\times R^{7}\times S^{2}$. Particularly, the $SO(3)$-isometry in
the transverse space implies that, as they approach, $D1$ and $\bar{D1}$ actually merge
and as a result a curvature singularity develops at the center $r=0$.  \\
Next we consider the $F1-\bar{F1}$ system given above in eq.(7)
and again take the limit $a\rightarrow 0$. 
\beqa
ds^2_{10} &=& \left(1 - {2m\over r}\right)[-dt^2 + dx^2_{1}] + \sum^{6}_{m=2} dx^2_{m} + dr^2
+ r^2\left(1 - {2m\over r}\right)(d\theta^2 + \sin^2 \theta d\phi^2), \nonumber \\ 
e^{2\phi} &=& \left(1 - {2m\over r}\right), ~~~B_{[2]} = 0.  
\eeqa 
In this limit, again it appears that the opposite electric $NSNS$ charges
carried by $F1$ and $\bar{F1}$ annihilate each other since $B_{[2]}=0$ and the metric solution
has the topology of $R\times R^{7}\times S^{2}$. Once again, the manifest $SO(3)$-isometry in
the transverse space implies that, as they are brought together, $F1$ and $\bar{F1}$ actually collide
and as a result a curvature singularity develops at the center $r=0$. \\
To summarize, as one can see in this supergravity descriptions, both $RR$ and $NSNS$-charged 
brane-antibrane systems are on equal footing in that they exhibit essentially the same semi-classical 
instabilities. And as the inter-brane distance gets smaller and smaller, say, towards the string 
scale $\sim \sqrt{\alpha'}=l_{s}$, we expect that these semi-classical instabilities should be taken over by the associated 
quantum, stringy instabilities. As we mentioned in the introduction, however, actually we have an
embarrassing state of affair since in the $NSNS$-charged case, the quantum entity that should take 
over the semi-classical instability at short length scales is missing. To be a little more concrete,
we invoke Sen's conjecture \cite{sen} for the decay/annihilation of unstable $D$-branes via the open 
string tachyon condensation. From the stringy perspective based on the open string field theory, 
Sen suggested that the eventual fate of the non-BPS $D_{p}-\bar{D}_{p}$ system could be a
supersymmetric vacuum via the open string tachyon condensation. Namely, the brane and the antibrane
merge and annihilate each other completely since firstly, the opposite $RR$ charges are cancelled
and secondly, the total energy of the system, upon merging, may vanish \cite{sen},
$E_{tot} = V(T_{0}) + 2M_{D} = 0.$ Thus according to this conjecture by Sen, the outcome of the
brane-antibrane collision could be a complete annihilation into a supersymmetric vacuum.
(We shall provide a detailed review of Sen's conjecture below in the next subsection.)
In the ``$NS$''-charged case, however, the situation changes. Namely, since
fundamental string does not end on another $F1$ nor on $NS5$ (namely no intersection rules such 
as $(0|F1, F1)$ or $(0|F1, NS5)$ exists), there is, as a result, no stringy description available 
for the brane-antibrane annihilation in terms of open string tachyon condensation via Sen's mechanism.
This absence of the quantum mechanism for the outcome of $F1-\bar{F1}$ or $NS5-\bar{NS5}$ annihilation
is indeed a very unnatural state of affair in light of the fact that 
$F1-\bar{F1}$ and $NS5-\bar{NS5}$ systems are just $U$ duals to $D_{p}-\bar{D}_{p}$ systems as we
have seen earlier. Certainly, therefore, a quantum, stringy description is in need
for these $F1-\bar{F1}$ and $NS5-\bar{NS5}$ annihilations into (presumably) supersymmetric vacua
and in the present work, we shall address this issue. 

\subsection{Stringy description of the instability at strong coupling}

Evidently, there is a combined gravitational and ($RR$) gauge attractions between the $RR$-charged
$D_{p}$ and $\bar{D}_{p}$ ($p=1,5$) at some large but finite separation leading to the
semi-classical instability. And it manifests itself in terms of the presence of conical
singularities owned by the supergravity solutions representing $D1-\bar{D1}$, $D5-\bar{D5}$ pairs
as mentioned above. As the inter-brane distance gets smaller and smaller toward, say, the distance
of order $a \sim \sqrt{\alpha'} = l_{s}$, this semi-classical description of the instability 
should be replaced by the quantum, stringy one that is represented by the tachyonic mode arising
in the spectrum of open strings stretched between $D_{p}$ and $\bar{D}_{p}$. Then as our supergravity
analysis given above indicates (in which the $NSNS$-charged ($F1-\bar{F1}$, $NS5-\bar{NS5}$) systems
are shown to be related, via the $S$-duality, to the $RR$-charged ($D1-\bar{D1}$, $D5-\bar{D5}$) 
systems), in a similar manner, the quantum, stringy description of the instabilities in the
($F1-\bar{F1}$, $NS5-\bar{NS5}$) systems should be related to that in the 
($D1-\bar{D1}$, $D5-\bar{D5}$) systems via the stringy (or brany) version of $S$-duality as well.
This natural anticipation thus leads us to propose the following quantum interpretation of the
instabilities that reside in the $NSNS$-charged brane-antibrane systems. Thus we first start with
the quantum description of the instabilities in the ($D1-\bar{D1}$, $D5-\bar{D5}$) systems. Now
consider the GSO projection for the open (super) string in which we use the convention that the 
$(-1)^{F}$ (``$F$'' here denotes the worldsheet fermion number carried by the string state)
eigenvalue of the $NS$ sector ground state is $-1$. Then from the observation that for the $NSNS$
sector, the closed string exchange interactions between the $D_{p}-D_{p}$ and between the 
$D_{p}-\bar{D}_{p}$ have the same sign whereas for the $RR$ sector, they have opposite sign, one
can deduce, using the closed string channel-open string channel duality, that the GSO projection
operator for strings in $D_{p}-D_{p}$ system is $(1/2)[1 + (-1)^{F}]$, while that for strings in
$D_{p}-\bar{D}_{p}$ system is $(1/2)[1 - (-1)^{F}]$. Namely, the open strings stretched between
$D_{p}$ and $\bar{D}_{p}$ have {\it wrong} GSO projection and thus develop tachyonic mode as
the lowest-lying state. This troublesome result of having the tachyonic mode in the non-BPS
$D_{p}-\bar{D}_{p}$ systems, however, has been circumvented by the idea of the unstable 
$D_{p}-\bar{D}_{p}$ system decay via the open string tachyon condensation. Particularly, according
to the conjecture due to Sen \cite{sen}, the condensation mechanism can be stated briefly as follows.
Consider, for example, a coincident $D_{p}-\bar{D}_{p}$ pair in type IIB theory.
Upon integrating out all the massive modes in the spectrum of open strings on the
$D_{p}-\bar{D}_{p}$ worldvolume one should get the tachyon potential and it gets its maximum 
at the false vacuum expectation value (vev) $T=0$. Next, since there is a $U(1)\times U(1)$ 
Born-Infeld gauge theory living in the worldvolume of this $D_{p}-\bar{D}_{p}$ system, 
the tachyon field $T$ picks up a phase under each of these $U(1)$ gauge transformations.
As a result, the tachyon potential $V(T)$ is a function only of $|T|$ and its minimum occurs 
at $T = T_{0}e^{i\theta}$ for some fixed true vev $T_{0}$. Then the essence of Sen's conjecture is
the proposition that at $T = T_{0}$, the sum of the tension of $D$-brane and anti-$D$-brane and the 
(negative) minimum negative potential energy of the tachyon is exactly zero
\beq
V(T_{0}) + 2M_{D} = 0 \nonumber
\eeq
with $M_{D}=T_{D}$ being the $D$-brane tension. And the philosophy behind this conjecture is the 
physical insight that the endpoint of the brane-antibrane annihilation would be a string vacuum 
in which the supersymmetry is fully restored. Thus this is the standard way (at least at the
moment) of describing the instabilities in the $RR$-charged brane-antibrane systems such as
($D1-\bar{D1}$, $D5-\bar{D5}$) pairs in quantum, stringy terms. \\
Consider now, its $S$-dual picture. Clearly, the $S$-dual of the $D1-\bar{D1}$ system with open
strings ($F1$) stretched between them should be the $F1-\bar{F1}$ system with $D$-strings ($D1$)
joining them while that of the $D5-\bar{D5}$ system with fundamental strings suspended between 
them should likewise be the $NS5-\bar{NS5}$ system with $D$-strings connecting them. (Indeed,
the $S$-dual of $(0|F1, D5)$, namely the intersection, $(0|D1, NS5)$ is actually known to
exist \cite{eric}.) The point on which we should be careful here is that, since the fundamental
open strings ($F1$) stretched between $D_{p}$ and $\bar{D}_{p}$ ($p=1, 5$) develop tachyonic modes
and hence are unstable, their $S$-dual partners, $D$-strings ($D1$) joining $F1$ and $\bar{F1}$
or $NS5$ and $\bar{NS5}$ should be {\it unstable} $D1$'s in IIB theory possessing again tachyonic 
modes. This is illustrated in Fig.1.
\begin{figure}[hbt]
\centerline{\epsfig{file=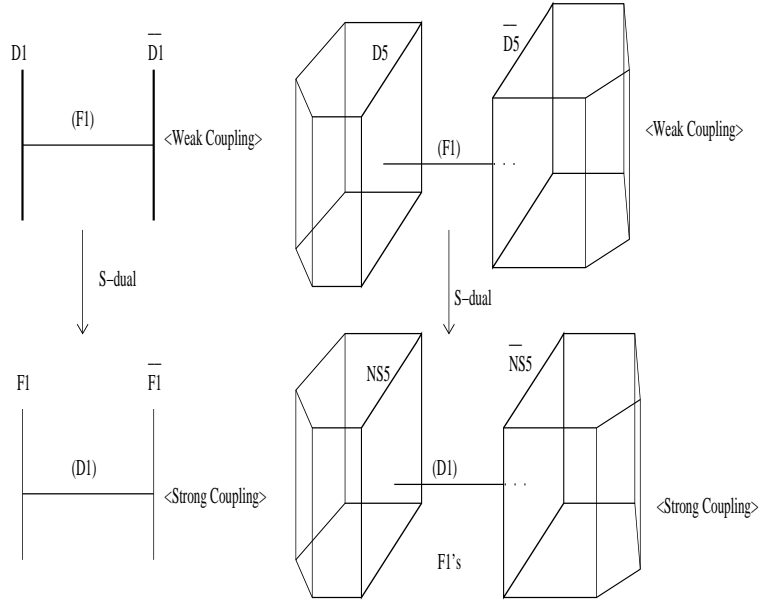, width=10cm, height=8cm}}
\caption{\ IIB theory ($F1-\bar{F1}$, $NS5-\bar{NS5}$) systems at strong coupling derived using
$S$-duality.}
\end{figure}
Along the line of Sen's conjecture then, one can anticipate that at the true vev $T=T_{0}$
of the tachyon on $D1$'s, the sum of the tensions of $NS$-branes and the (negative) minimum
potential energy of the tachyon should exactly be zero, $V(T_{0})+2M_{NS}=0$,
which is just the $S$-dual of the eq.(12) given earlier.
Indeed, particularly the fact that the $D1$'s stretched between IIB $NS5$ and $\bar{NS5}$ possesses 
tachyonic mode and hence are responsible for the quantum instability of this system 
has been pointed out earlier in the literature invoking the self $S$-dual nature of 
IIB theory \cite{lozano2} just as discussed above or using the fact \cite {lozano1} that
this IIB system is the $T$-dual of the IIA system of $D2$'s stretched between IIA $NS5$ and 
$\bar{NS5}$ whose quantum instability, in turn, is inherited from that of the M-theory system 
of $M2$'s stretched between $M5$ and $\bar{M5}$ \cite{yi}. Thus a new ingredient in the present 
study is that this quantum description has been extended to the $F1-\bar{F1}$ pair as well in which 
$D1$'s stretched between them are again responsible, via their tachyonic modes, for the quantum 
instabilities.

\subsection{Stringy description of the instability at weak coupling}

The discussion so far, however, is valid only at strong coupling.
Namely note that we have derived the stringy (or brany) description for the instabilities in the 
$NSNS$-charged brane-antibrane systems in this IIB theory using basically the $S$-duality by 
which the weak coupling picture of instabilities in $RR$-charged brane-antibrane systems have been 
mirrored to the strong coupling picture of those in $NSNS$-charged ones. As such, strictly speaking,
this stringy (or brany) description for the instabilities in the $NSNS$-charged brane-antibrane 
systems holds true in the strong coupling regime. And since the systems are not BPS, the 
extrapolation of this particular description to the weak coupling regime would not be safe, 
and more careful treatment there is required. Thus in the following, we attempt at providing the 
origin of quantum instabilities of the same IIB systems at weak coupling. To this end,
first notice that the tensions of fundamental string ($F1$) and $D$-string ($D1$) are given
respectively by $T_{F1} = 1/(2\pi \alpha')$, $T_{D1} = 1/(2\pi \alpha' g_{s})$ with 
$\alpha' = l^2_{s}$ and $g_{s}$ being the string length squared and the string coupling 
respectively. At weak coupling $g_{s}\rightarrow 0$, which is the familiar case, the $D$-string
gets heavy and the fundamental string is the lightest object in the theory and hence one
should identify the $D$-string instability (as that of $D1-\bar{D1}$ system) as originating from 
the tachyonic modes of the fundamental strings ending on the $D$-string. And the weak coupling
description of the unstable $D1-\bar{D1}$ system depicted in Fig.1 given earlier exhibits such a
situation. At strong coupling $g_{s}\rightarrow \infty$, however, now the fundamental string is 
heavy and the $D$-string becomes the lightest object in the theory. Thus the vibrating modes of 
the $D$-string can be seen in perturbation theory and this is why one can identify the $D$-string 
instability in terms of its {\it own} lowest-lying tachyonic mode as discussed above and in the 
literature \cite{lozano1,lozano2}. Now, precisely this last point provides the clue to our main
question here as to the origin of quantum instabilities of type IIB $NSNS$-charged brane-antibrane
systems at weak coupling. Namely as discussed above and shown in Fig.1 (the strong coupling side), 
at stong coupling, it is the tachyonic mode of the {\it light} $D$-string itself that is responsible 
for the quantum instability of the unstable, {\it heavy} fundamental string (such as that of 
$F1-\bar{F1}$ system). Now, consider the ($D1-\bar{D1}$, $D5-\bar{D5}$) systems at {\it strong} 
coupling. Then the first thing that would come to one's mind is that their quantum instabilities 
might come from the $F$-strings stretched between them. For reasons just stated, however, this is 
not really the case as the $F$-string is now heavy and hence does not admit perturbative description 
for its vibrational modes. Instead, the instabilities essentially come from the tachyonic modes of 
light $D$-strings ending on this heavy $F$-string. And of course this would be true because now the 
$D$-string is the lightest object in the theory and it does have intersection with the $F$-string, 
i.e., $(0|F1,D1)$. This realization of the role played by the light $D$-strings 
(ending on the heavy $F$-string generally stretched between $D_{p}$ and $\bar{D}_{p}$) 
as an actual source of quantum instabilities at strong coupling is what distinguishes the situation 
for the case at hand from what happens in the familiar weak coupling case discussed in the previous
subsection and in the literature \cite{lozano1,lozano2}. 
Finally, then, the origin of instabilities in IIB theory ($F1-\bar{F1}$, $NS5-\bar{NS5}$) systems at 
{\it weak} coupling can be deduced from that of ($D1-\bar{D1}$, $D5-\bar{D5}$) systems at {\it strong} 
coupling just discussed again via $S$-duality. Namely under $S$-duality, this picture of unstable
($D1-\bar{D1}$, $D5-\bar{D5}$) systems at strong coupling is mapped into 
the IIB theory ($F1-\bar{F1}$, $NS5-\bar{NS5}$) systems at weak coupling in which {\it heavy} 
$D$-string stretched between ($F1-\bar{F1}$, $NS5-\bar{NS5}$) has light $F$-strings ending on it 
whose lowest-lying tachyonic modes ultimately represent the quantum instabilities of these 
$NSNS$-chaged  brane-antibrane systems. And clearly this is consistent with the familiar weak coupling 
description discussed earlier in which the tachyonic mode of the fundamental string is responsible
for the quantum instability of the heavy $D$-string (and of other non-BPS $D_{p}$-branes as well).  
This suggestion for the origin of quantum instabilities of the $NSNS$-charged 
brane-antibrane systems at weak coupling has been depicted in Fig.2. 
\begin{figure}[hbt]
\centerline{\epsfig{file=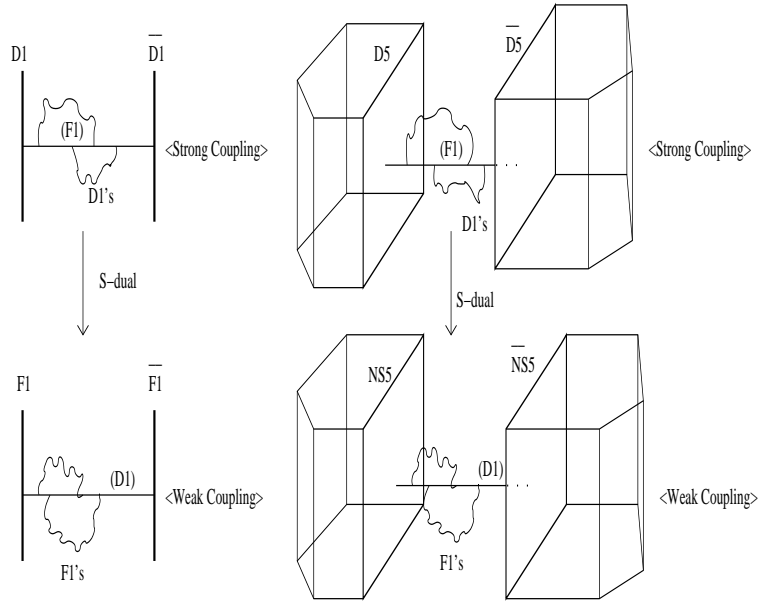, width=10cm, height=8cm}}
\caption{\ IIB theory ($F1-\bar{F1}$, $NS5-\bar{NS5}$) systems at weak coupling derived using
$S$-duality.}
\end{figure}
Therefore for the $RR$-charged brane-antibrane systems at strong coupling, one should expect
\beq
\tilde{V}(\tilde{T}_{0}) + 2M_{D1} + M_{F1} = 0,
~~~\tilde{V}(\tilde{T}_{0}) + 2M_{D5} + {1\over \alpha'^2}M_{F1} = 0
\eeq
for $D1-\bar{D1}$, $D5-\bar{D5}$ systems respectively
and for their $S$-dual, $NSNS$-charged brane-antibrane systems at weak coupling, one should expect
\beq
\tilde{V}(\tilde{T}_{0}) + 2M_{F1} + M_{D1} = 0,
~~~\tilde{V}(\tilde{T}_{0}) + 2M_{NS5} + {1\over \alpha'^2}M_{D1} = 0
\eeq
for $F1-\bar{F1}$, $NS5-\bar{NS5}$ systems respectively and
where $M_{F1} = T_{F1} = 1/2\pi \alpha'$ and $M_{NS5} = T_{NS5} = 1/(2\pi)^{5}\alpha'^{3}g^2_{s}$.
Note that the last term, the $F$ ($D$)-string tension $M_{F1}=T_{F1}$
($M_{D1}=T_{D1}$) in eq.(13 (14)) (up to the factor ${1/ \alpha'^2}$, being
introduced for dimensional consideration),
which was negligible at weak (strong) coupling, is now no longer so. 
$\tilde{V}(\tilde{T}_{0})$ is the (negative) minimum of potential of the tachyon arising in the 
spectrum of open $D$ ($F$)-strings ending on unstable
$F$ ($D$)-strings. Obviously, it is determined separately for $D1-\bar{D1}$ ($F1-\bar{F1}$) and 
$D5-\bar{D5}$ ($NS5-\bar{NS5}$) pairs and should be different in its structure and hence in value 
from $V(T_{0})$ given in eq.(12) in the $RR$-charged brane-antibrane systems at weak coupling. 
And this is our proposal for the quantum, stringy description for the instabilities in the 
$NSNS$-charged ($F1-\bar{F1}$, $NS5-\bar{NS5}$) systems at {\it weak} coupling in IIB theory that 
has remained unexplained thus far. Lastly, it is worthy of note that there is an important 
lesson to be learned from this study of instabilities in the IIB theory $RR$-charged
brane-antibrane systems at weak string coupling. It is the fact that the original Sen's
conjecture for the $RR$-charged brane-antibrane annihilation via the tachyon condensation 
represented by $V(T_{0}) + 2M_{D} = 0$ given above has its relevance only at {\it weak} coupling. 
And at {\it strong} coupling, for which the fundamental open strings stretched between $D_{p}$
and $\bar{D}_{p}$ in IIA/IIB theories becomes relatively heavy and hence should have much
lighter $D$-strings ending on them as an indicator for their quantum instability, the corresponding 
equation representing Sen's conjecture should be given instead by
the ones in eq.(13) with the tachyon arising in the spectrum of open
$D$-string instead. Thus in this way we extended Sen's conjecture at weak 
coupling to strong coupling as well.

\section{Instabilities in IIA theory ($F1-\bar{F1}$, $NS5-\bar{NS5}$) systems}\label{ }

Since the $NSNS$-charged $F1$ and $NS5$ arise in the spectrum
of both IIA and IIB theories, the analysis of instabilities would be incomplete unless we discuss
the IIA theory case as well. Thus we now turn to the discussion of the instabilities in the
($F1-\bar{F1}$, $NS5-\bar{NS5}$) systems in type IIA theory. 

\subsection{Supergravity description of the instability}

For the present case of IIA theory, since the relevant string duality to employ is the M/IIA
duality, in the following we demonstrate that starting with the unstable M-theory
brane-antibrane solutions, one can perform appropriate KK-reductions along the M-theory
circle direction to obtain non-BPS $F1-\bar{F1}$ and $NS5-\bar{NS5}$ solutions. Now starting
with the supergravity solution representing the system of $M2-\bar{M2}$ pair in $D=11$ \cite{kim},
\beqa
ds^2_{11} &=& H^{-2/3}[-dt^2 + \sum^{2}_{i=1}dx^2_{i}] + H^{1/3}[\sum^{7}_{m=3}dx^2_{m} + 
(\Delta+a^2\sin^2 \theta)\left({dr^2\over \Delta}+d\theta^2\right) + \Delta \sin^2 \theta d\phi^2], 
\nonumber \\
A^{11}_{[3]} &=& -\left[{2ma\cos \theta \over \Sigma}\right]dt\wedge dx_{1}\wedge dx_{2}
\eeqa
again with the ``modified'' harmonic function given by 
$H(r) = \Sigma /({\Delta + a^2\sin^2 \theta})$ and performing the KK-reduction along a direction
longitudinal to the $M2~(\bar{M2})$ brane worldvolume, $x_{2}$,
\beqa
ds^2_{11} &=& e^{-{2\over 3}\phi}ds^2_{10} + e^{{4\over 3}\phi}(dx_{2}+A_{\mu}dx^{\mu})^2, 
\nonumber \\
A^{11}_{[3]} &=& A^{IIA}_{[3]} + B_{[2]}\wedge dx_{2}
\eeqa
one can obtain the supergravity solution representing the system of $F1-\bar{F1}$ pair given 
earlier in eq.(7) in the previous section. 
Next, starting with the $M5-\bar{M5}$ solution in $D=11$ \cite{kim},
\beqa
ds^2_{11} &=& H^{-1/3}[-dt^2 + \sum^{5}_{i=1}dx^2_{i}] + H^{2/3}[dx^2_{6} + dx^2_{7} + 
(\Delta+a^2\sin^2 \theta)\left({dr^2\over \Delta}+d\theta^2\right) + \Delta \sin^2 \theta d\phi^2], 
\nonumber \\
F^{11}_{[4]} &=& {{2mra\sin^2 \theta (r^2 + a^2\cos^2 \theta)}\over {(\Delta + a^2 \sin^2 \theta)^2}}
dx_{6}\wedge dx_{7}\wedge dr\wedge d\phi \\
&-& {{4mra\sin \theta \cos \theta \Delta}\over {(\Delta + a^2 \sin^2 \theta)^2}}
dx_{6}\wedge dx_{7}\wedge d\theta \wedge d\phi \nonumber 
\eeqa
where again $H(r) = \Sigma /({\Delta + a^2\sin^2 \theta})$ and carrying out the KK-reduction along 
a direction transverse to the $M5~(\bar{M5})$ brane worldvolume, $x_{7}$,
\beqa
ds^2_{11} &=& e^{-{2\over 3}\phi}ds^2_{10} + e^{{4\over 3}\phi}(dx_{7}+A_{\mu}dx^{\mu})^2, 
\nonumber \\
F^{11}_{[4]} &=& F^{IIA}_{[4]} + H^{m}_{[3]}\wedge dx_{7}
\eeqa
one can arrive at the $NS5-\bar{NS5}$ solution with $H^{m}_{[3]}=dB^{m}_{[2]}$ given earlier
in eq.(9).  Again, the fact that these M-theory brane-antibrane pair or the IIA theory $F1-\bar{F1}$
and $NS5-\bar{NS5}$ pairs represented by the supergravity solutions given above indeed exhibit
semi-classical instabilities in terms of the appearance of the conical singularities
can be found in detail in our earlier work \cite{kim}. 

\subsection{Stringy description of the instability at strong coupling}

We now move on to the quantum, stringy perspective. In order eventually to identify the quantum
entity that is supposed to take over the semi-classical instability as the inter-brane distance
gets smaller, say, toward the string scale $\sim \sqrt{\alpha'} = l_{s}$, in the 
($F1-\bar{F1}$, $NS5-\bar{NS5}$) systems in type IIA theory, we start with their M-theory
counterparts. Recall first that the $D=11$ M-theory possesses 5-types of brane solutions ;
\beq
M-{\rm wave}, ~~~M2, ~~~M5, ~~~MKK-{\rm monopole}, ~~~M9 \nonumber
\eeq
and particularly the intersection rules known among $M2$ and $M5$ branes are given by \cite{eric} ;
\beq
(0|M2, M2), ~~~(1|M2, M5), ~~~(1|M5, M5), ~~~(3|M5, M5).  \nonumber
\eeq
Thus from these intersection rules, we can deduce the following ``triple'' intersections.  
(Henceforth $M5-(M2)-\bar{M5}$, for example, indicates the configuration in which $(M2)$ is
stretched between $M5$ and $\bar{M5}$.)

{\bf \rm (I) $[M2-(M2)-\bar{M2}]$ $\rightarrow$ $[F1-(D2)-\bar{F1}]$}
$$
\begin{array}{lc|cccccccccc}
M2: & 0 & 1 & 2 & - & - & - & - & - & - & - & - \\
\bar{M2}: & 0 & 1 & 2 & - & - & - & - & - & - & - & - \\
(M2): & 0 & -  & - & 3 & 4 & - & - & - & - & - & -
\end{array}
$$
where we used $(0|M2, M2)$ and $(0|M2, \bar{M2})$. Using now the M/IIA-duality, consider the
KK-reduction following the compactification along the direction $x_{2}$ longitudinal both
to $M2$ and $\bar{M2}$ but transverse to $(M2)$ worldvolume to get
$$
\begin{array}{lc|cccccccccc}
F1: & 0 & 1 & - & - & - & - & - & - & - & - \\
\bar{F1}: & 0 & 1 & - & - & - & - & - & - & - & - \\
(D2): & 0 & -  & 2 & 3 & - & - & - & - & - & -
\end{array}
$$
which is consistent with the known intersection rule $(0|F1, D2)$.

{\bf \rm (II) $[M5-(M2)-\bar{M5}]$ $\rightarrow$ $[NS5-(D2)-\bar{NS5}]$}
$$
\begin{array}{lc|cccccccccc}
M5: & 0 & 1 & 2 & 3 & 4 & 5 & - & - & - & - & - \\
\bar{M5}: & 0 & 1 & 2 & 3 & 4 & 5 & - & - & - & - & - \\
(M2): & 0 & 1  & - & - & - & - & 6 & - & - & - & -
\end{array}
$$
where we used $(1|M2, M5)$ and $(1|M2, \bar{M5})$.
In this time, we consider the compactification followed by the KK-reduction along the direction
$x_{7}$ transverse both to $M5(\bar{M5})$ and $(M2)$ worldvolumes to arrive at
$$
\begin{array}{lc|cccccccccc}
NS5: & 0 & 1 & 2 & 3 & 4 & 5 & - & - & - & - \\
\bar{NS5}: & 0 & 1 & 2 & 3 & 4 & 5 & - & - & - & - \\
(D2): & 0 & 1  & - & - & - & - & 6 & - & - & -
\end{array}
$$
which is consistent with the known intersection rule $(1|D2, NS5)$ \cite{eric}. 
Note here that for this 
case (II), one may instead consider the KK-reduction along the direction $x_{6}$ which is transverse
to $M5(\bar{M5})$ but longitudinal to the $(M2)$ worldvolume to get $NS5-(F1)-\bar{NS5}$
consistently with the known intersection rule $(1|F1, NS5)$ \cite{eric}. 
This last option, however, is
irrelevant to the present discussion of the quantum interpretation of the instability in
$NS5-\bar{NS5}$ system since $F1$ here is embedded in the $NS5$ and $\bar{NS5}$ entirely.
Now since the M-theory
membrane ($(M2)$) stretched between $M2$ and $\bar{M2}$ or $M5$ and $\bar{M5}$ should represent
quantum instability presumably in terms of ``string-like'' tachyonic modes arising on it as suggested 
by Yi \cite{yi}, their M/IIA-dual partners, i.e., the $(D2)$-branes connecting $F1$ and $\bar{F1}$
or $NS5$ and $\bar{NS5}$ should be {\it unstable} $D2$'s in IIA theory possessing again tachyonic 
modes. It is also worthy of note that for the case (I), the KK-reduction along the direction, say, 
$x_{3}$ transverse to $M2(\bar{M2})$ but longitudinal to $(M2)$ would yield $D2-(F1)-\bar{D2}$, while
for the case (II), the KK-reduction along the direction $x_{1}$ longitudinal both to $M5(\bar{M5})$ 
and $(M2)$ would lead to $D4-(F1)-\bar{D4}$. Clearly, these are the familiar $RR$-charged 
brane-antibrane systems in type IIA theory in which the quantum, stringy description of the
instabilties is given in terms of the open string tachyon condensation in the sense of Sen's
conjecture. And it is precisely this point that provides the logical ground on which one can argue
that the $(M2)$-brane stretched between $M2$ and $\bar{M2}$ or $M5$ and $\bar{M5}$ would develop
``string-like'' tachyonic modes (or tachyonic strings) and hence represent quantum instabilities
in these M-theory brane-antibrane systems. Then again along the line of Sen's conjecture, one 
immediately anticipates $\bar{V}(\bar{T}_{0})+2M_{NS}=0$.
As mentioned earlier when we discussed the type IIB-theory case, the argument 
particularly on the origin of instabilities in IIA theory $NS5-\bar{NS5}$ system
precisely of this sort has been pointed out in the 
literature \cite{lozano1,lozano2}. Thus again a new ingredient in the present 
work is that this quantum description has been extended to the $F1-\bar{F1}$ pair as well
(see Fig.3.).
\begin{figure}[hbt]
\centerline{\epsfig{file=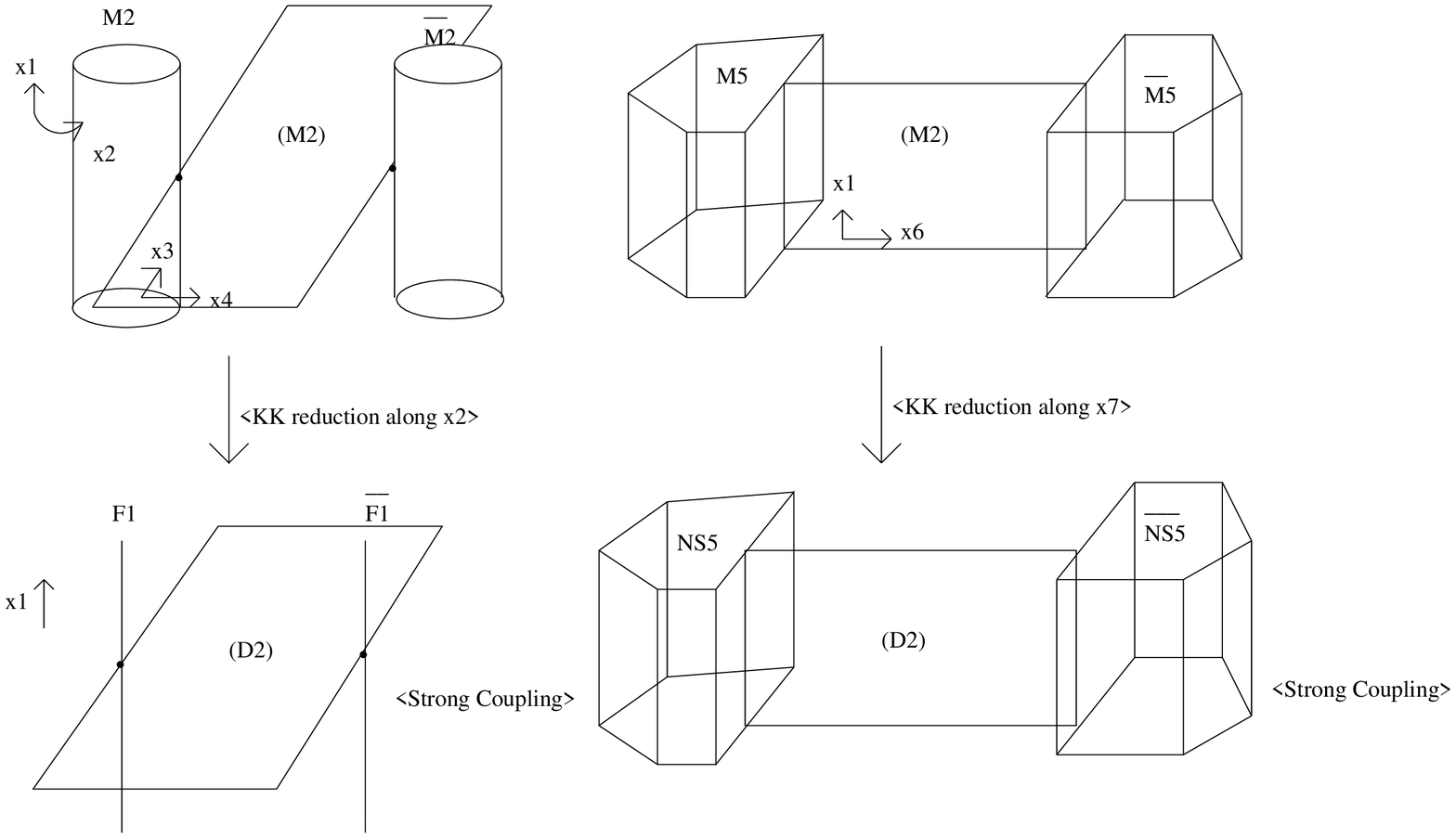, width=10cm, height=8cm}}
\caption{\ IIA theory ($F1-\bar{F1}$, $NS5-\bar{NS5}$) systems at strong coupling derived using
M/IIA-duality.}
\end{figure}

\subsection{Stringy description of the instability at weak coupling}

The problem is, however, that the discussion thus far is again valid only at strong 
coupling and cannot be safely extrapolated to the weak coupling regime since the systems under
consideration are not BPS. 
Namely note that we have derived the stringy (or brany) description for the instabilities 
in the $NSNS$-charged brane-antibrane systems in IIA theory using basically the the M/IIA-duality 
by which the instabilities in the M-theory brane-antibrane systems have been dualized
to the strong coupling picture of their 10-dimensional $NSNS$-charged counterparts. 
As before therefore, in the following we attempt to provide the origin of quantum instabilities of 
the IIA $NSNS$ brane-antibrane systems at weak string coupling. Unlike the strong coupling case
discussed above, however, here we cannot employ the M/IIA-duality as there is no M-theory
dual for weak coupling limit of type IIA string theory. Thus we should now consider some
other way and it turns out that the origin of quantum instabilities of the IIA $NSNS$ brane-antibrane 
systems at weak string coupling can be unveiled by taking the $T$-dual of its IIB theory
counterpart discussed in sect.2.3. Namely, start with our earlier study of the origin of quantum 
instabilities of IIB $NSNS$ brane-antibrane systems at weak coupling. First, for the IIB
$F1-\bar{F1}$ case, in which the $D$-string stretched between $F1$ and $\bar{F1}$ is heavy and 
hence the tachyonic modes of the light $F$-strings ending on it are responsible for quantum 
instabilities, we take a $T$-dual of this system along a direction, say $x_{3}$, which is
transverse to all $F1$'s, $\bar{F1}$ and $D1$. Then one ends up with the system in which now
{\it heavy} $D2$ stretched between $F1$ and $\bar{F1}$ has light $F$-strings ending on it whose
lowest-lying tachyonic modes represent the quantum instability of this system. And this is because
under this particular $T$-dual, all $F1$'s and $\bar{F1}$ are mapped to themselves.
Next, for the IIB $NS5-\bar{NS5}$ case, take a $T$-dual along a direction, say $x_{1}$, which is
transverse to all $F1$'s and $D1$ but longitudinal to $NS5$ and $\bar{NS5}$. In this way, one is
left with the system in which {\it heavy} $D2$ stretched between $NS5$ and $\bar{NS5}$ has light 
$F$-strings ending on it whose lowest-lying tachyonic modes again represent the quantum instability 
of the system. And this is because under this particular $T$-dual, all $F1$'s and $NS5$ and 
$\bar{NS5}$ are mapped to themselves. This suggestion has been depicted in Fig.4. 
\begin{figure}[hbt]
\centerline{\epsfig{file=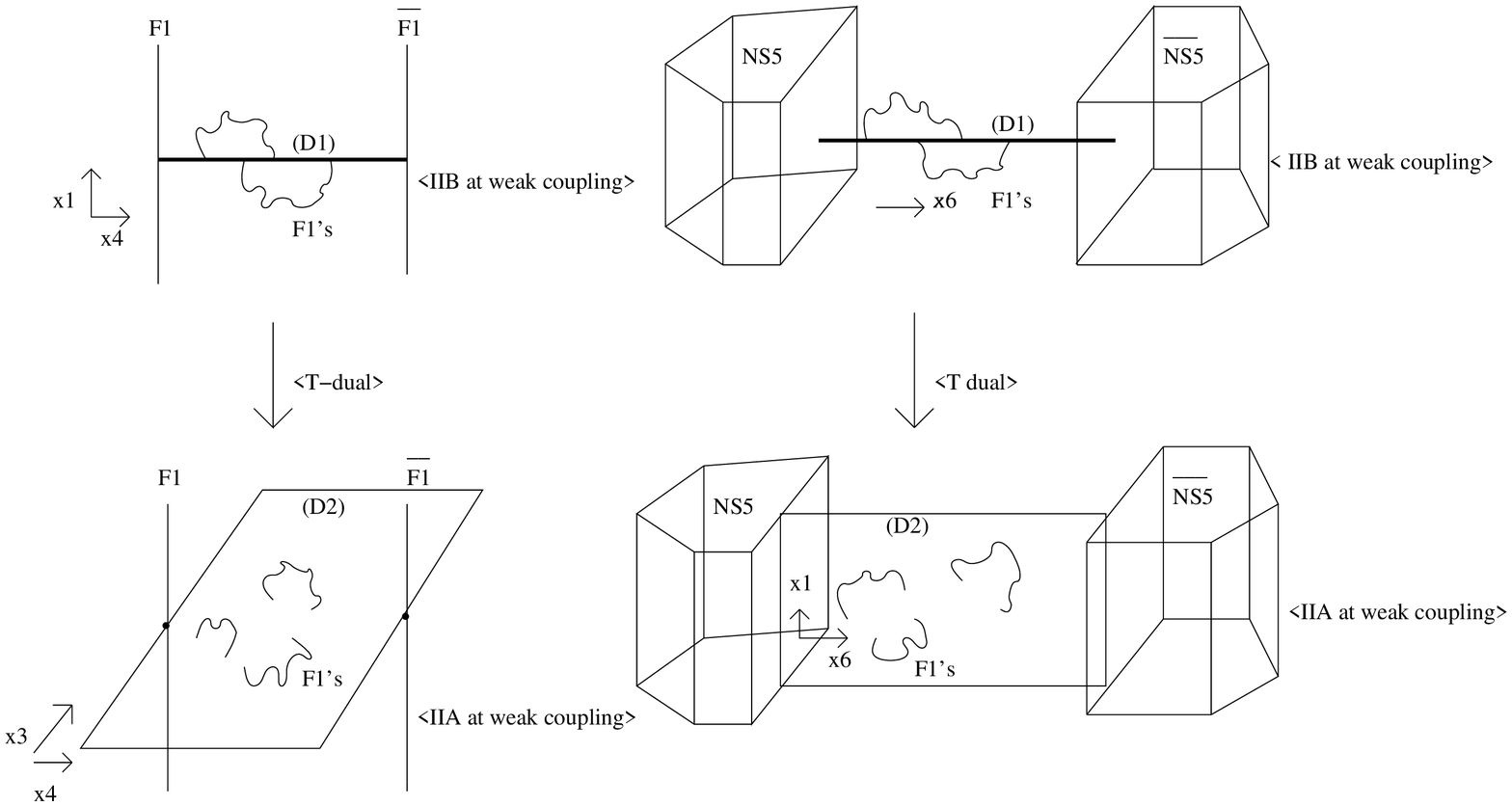, width=10cm, height=8cm}}
\caption{\ IIA theory ($F1-\bar{F1}$, $NS5-\bar{NS5}$) systems at weak coupling derived using
$T$-duality.}
\end{figure}
Therefore at weak coupling, one should expect instead 
\beq
\bar{V}(\bar{T}_{0}) + 2M_{F1} + {\alpha'^{1/2}}M_{D2} = 0,
~~~\bar{V}(\bar{T}_{0}) + 2M_{NS5} + {1\over \alpha'^{3/2}}M_{D2} = 0
\eeq
which are the $T$-dual versions of eq.(14).
Note that the last terms, the $D2$-brane tension $M_{D2}=T_{D2}=1/(2\pi)^{2}\alpha'^{3/2} g_{s}$,
(again up to the $\alpha'$-dependent factors being introduced for dimensional reason)
which was negligible at strong coupling, is now no longer so. 
$\bar{V}(\bar{T}_{0})$ is the (negative) minimum of potential of the tachyon arising in the 
spectrum of open {\it fundamental} strings ending on unstable $D2$-branes. 
Apparently, it is determined separately for $F1-\bar{F1}$ and $NS5-\bar{NS5}$ pairs
and should be different from its IIB theory counterpart $\tilde{V}(\tilde{T}_{0})$ 
given earlier in eq.(14) or from $V(T_{0})$ appearing in the $RR$-charged brane-antibrane systems 
given in eq.(12). 
Before we leave this subsection, we comment on yet another possibility. 
That is, starting with the M-brane intersection rule $(1|M2, M-wave)$ \cite{eric}, one can, 
via the KK-reduction along the direction longitudinal both to $M-wave$ and $M2$, the intersection 
$(0|F1,D0)$, which, in turn, implies the $F1-(D0)-\bar{F1}$ situation. 
Thus at strong coupling, it would be the tachyonic mode of 
the light $D0$ itself while at weak coupling, it would be that of the light $F$-strings ending on 
the heavy $D0$, that could also be responsible for the quantum instability of IIA theory $F1-\bar{F1}$
system in addition to the situation associated with the $D2$ discussed above. Analogous situation
associated with IIA theory $NS5-\bar{NS5}$ system, namely the configuration $NS5-(D0)-\bar{NS5}$,
however, does not arise. This is because the potentially relevant intersection rule 
$(1|M5, M-wave)$ \cite{eric},
leads, via the KK-reduction along the direction transverse both to $M-wave$ and $M5$, to 
the intersection $(1|NS5,W)$, (with $W$ denoting the IIA theory $pp$-wave) which implies the 
$NS5-(W)-\bar{NS5}$ situation. Evidently, this configuration is not relevant for the present 
discussion as the IIA theory $pp$-wave, which is entirely embedded in $NS5$ (and $\bar{NS5}$) is
not known to be tachyonic. 
And this is our proposal for the quantum, stringy description for the instabilities in the 
$NSNS$-charged ($F1-\bar{F1}$, $NS5-\bar{NS5}$) systems at {\it weak} coupling in IIA theory that 
has not been discussed thus far.

\section{Concluding remarks}\label{ }

Firstly, it is worthy of note that there is an important lesson we learned in the present study of
quantum instabilities of $RR$ and $NSNS$-charged brane-antibrane systems.
That is, at {\it strong} coupling, the fundamental open strings stretched between 
$RR$-charged $D_{p}$ and $\bar{D}_{p}$ in IIA/IIB theories become relatively heavy and hence should 
have much lighter $D$-strings ending on them as an indicator for their quantum instability.
And via the $S$-dual of this picture, we then realised that at {\it weak} coupling, the $D$-strings 
stretched between $NSNS$-charged $F1-\bar{F1}$ and $NS5-\bar{NS5}$ systems in IIA/IIB theories become 
relatively heavy and hence should likewise have much lighter fundamental strings ending on them again as 
an indicator for their quantum instability. In this realisation, however, one might wonder if these
$F1's$ and $D1's$ might form bound states and then add further instability to the system.
In some intermediate coupling regime, there indeed may be possibility of forming such $F1$-$D1$
bound states that would spoil the validity of our suggestion stated above.  To see this in some
more detail, we follow the argument of Witten \cite{witten}. Consider the system of, say, a
$F$-string and a $D$-string in parallel. The total tension of this system reads
$T_{F1}+T_{D1} = (g_{s}+1)/(2\pi \alpha' g_{s})$ which is certainly greater that that of a 
$F1-D1$ BPS bound state \cite{witten}, $T_{F1-D1}= \sqrt{g^2_{s}+1}/(2\pi \alpha' g_{s})$ for some
finite, intermediate value of the coupling $g_{s}$. Thus this parallel configuration is not
supersymmetric but it can lower its energy till it reaches the the BPS bound above and become
eventually supersymmetric. Namely, the $F$-string breaks, its endpoints being attached to the
$D$-string (or more generally one ends on the other) and then moves off to infinity. Meanwhile, 
$F$-string endpoints are charged under the $D$-string gauge field. Therefore, if this happens, 
a flux runs between the endpoints and hence the final configuration is a $D$-string with a parallel
flux or in effect, a $D$-string with the $F$-string dissolved on it. And a detailed 
calculation \cite{witten} shows that the final tension then saturates the BPS bound. All this
argument on the possibility of forming a $F1-D1$ bound state, however, is relevant at some
intermediate coupling. At either very strong or 
very weak coupling regime, $F1's$ and $D1's$ have nearly negligible chance to form such 
bound states and hence this kind of situation becomes irrelevant to consider. To see this explicitly,
consider again the system of $F1$ and $D1$ (i.e., one ending on the other) with the total tension 
being given by $T_{F1}+T_{D1} = 1/(2\pi \alpha') + 1/(2\pi \alpha' g_{s})$. At very strong
coupling, it reduces just to that of $F1$, i.e., $T_{F1} = 1/(2\pi \alpha')$ whereas at very weak
coupling it becomes almost that of $D1$, i.e., $T_{D1} = 1/(2\pi \alpha' g_{s})$. This means that
either at very strong or at very weak coupling, the two cannot behave as a bound state but instead 
one is very light and provides a vibrational spectrum, particularly the tachyonic mode, while ending on 
the other heavy string. And certainly, we do not call this a bound state of two strings of 
comparable masses (tensions, for the case at hand). \\  
Next, it seems relevant to check if the quantum description for the instabilities in the 
$NSNS$-charged brane-antibrane systems in IIA/IIB theories discussed in the present work is
indeed consistent with some known wisdom as of now. Thus we particularly focus on the case of
$NS5-\bar{NS5}$ systems in IIA/IIB theories. The nature of $NS5$-brane worldvolume dynamics 
uncovered thus far \cite{ns5} can be summarized as follows. 
Using the worldvolume dynamics of $RR$-charged
$D$-branes and invoking the string dualities, it has been derived that the non-chiral type IIA
string theory gives rise to a chiral $NS5$-brane worldvolume theory with $(2,0)$ supersymmetry
in 6-dimensions while the chiral type IIB string theory yields a non-chiral $NS5$-brane with
$(1,1)$ supersymmetry. Thus the light fields on a single IIA $NS5$-brane worldvolume correspond
to a tensor multiplet of 6-dimensional (2,0) SUSY, consisting of a self-dual $B_{\mu\nu}$ field
and 5-scalars (and fermions). Meanwhile on a single IIB $NS5$-brane worldvolume, there is a vector
multiplet consisting of a 6-dimensional gauge field and 4-scalars (and fermions). The four of the
5-scalars in the tensor multiplet on the IIA $NS5$-brane and the 4-scalars in the vector
multiplet on the IIB $NS5$-brane describe fluctuations of the $NS5$-brane in the transverse
directions. Next, the low energy worldvolume dynamics on a stack of $N$-coincident IIB $NS5$-branes
is a 6-dimensional $(1,1)$ $SU(N)$ SYM theory arising from the ground states of $D$-strings 
stretched between them. And the low energy worldvolume theory on a pile of $N$-coincident IIA
$NS5$-branes is a non-trivial field theory with $(2,0)$ SUSY in 6-dimensions. Particularly, it
contains string-like low energy excitations corresponding to $D2$-branes stretched between them.
Certainly this last point, namely that $D1$-branes and $D2$-branes provide the low energy
worldvolume dynamics on the $NS5$-branes in type IIB and IIA theories respectively, is indeed
consistent with our findings in the present work. That is, unstable $D1$ and unstable $D2$ each 
with (at weak coupling) or without (at strong coupling)
fundamental strings ending on each of them possessing tachyonic modes, are responsible for the
quantum instabilities in the $NS5-\bar{NS5}$ pairs in type IIB and IIA theories respectively.
Next, to repeat, the main purpose of the present work was to uncover concrete quantum interpretation of the 
instabilities in the $NSNS$-charged brane-antibrane systems in type IIA/IIB theories in terms
of stringy description. And we were particularly interested in the case when the endpoint of
the brane-antibrane annihilation is the string vacuum in which the supersymmetry is fully
restored. Now, going back to the instabilities in the $RR$-charged brane-antibrane systems,
there is, again according to Sen's conjecture \cite{sen}, the other channel for the decay of 
$D_{p}-\bar{D}_{p}$ systems in which the supersymmetry is partially restored and there the
endpoint is a stable $D_{(p-2)}$-brane realised as a topological soliton. To be a little
more specific, starting with a non-BPS $D_{p}-\bar{D}_{p}$ system in type IIA/IIB theories,
in the next step a non-BPS $D_{(p-1)}$-brane may result as the tachyonic kink on the
brane-antibrane pair and then in the last stage a BPS $D_{(p-2)}$-brane can emerge again as
the tachyonic kink on the non-BPS $D_{(p-1)}$-brane worldvolume. In this spirit, the
$NSNS$-charged $NS5-\bar{NS5}$ system in IIB theory may settle down to a BPS $D3$-brane since
in this self $S$-dual IIB theory, $NS5-\bar{NS5}$ system is $S$-dual to $D5-\bar{D5}$ system 
and $D3$ is $S$-dual to itself, as has been pointed out first by Yi \cite{yi}. The decay of
the $NS5-\bar{NS5}$ system possibly to another lower-dimensional brane in IIA theory, on the
other hand, should be treated in a rather different fashion as the $NS5-\bar{NS5}$ pair in
IIA theory does not have any link to $RR$-charged $D-\bar{D}$ pair. Here, the relevant string
duality one can turn to is the M/IIA-duality. Thus from this point on, Yi \cite{yi} also 
argued that the $NS5-\bar{NS5}$ pair in IIA theory may relax to a $D2$-brane as it can be deduced
from the $S^1$ compactification of a M-theory picture in which $M5$ and $\bar{M5}$ annihilate 
into a $M2$. One can quantify this argument as follows. Consider the Chern-Simons term in
the effective worldvolume theory action for $NS5-\bar{NS5}$ system which would essentially be
the same as that for $M5-\bar{M5}$ system given by
\beq
\int_{R^{5+1}}C_{[3]}\wedge H_{[3]}
\eeq
where $C_{[3]}$ and $H_{[3]}$ are the descendants (to $D=10$) of 3-form tensor potential 
of $D=11$ supergravity and the field strength of the 2-form living in the worldvolume of $M5$-brane 
\cite{yi,lozano1} respectively. Now integrating over a localized and quantized magnetic flux
$H_{[3]}$ on a transverse $R^3$, one ends up with
\beq
\int_{R^{2+1}}C_{[3]}
\eeq
which is just the way how $C_{[3]}$ would couple to a $D2$-brane implying that the localized and
quantized magnetic $H$ flux should be identified with a $D2$-brane.   
Note that this mechanism by which one can identify the $D2$-brane as a topological soliton
emerging from the tachyon condensation via the (non) abelian Higgs mechanism works both at
strong and at weak couplings. But what distinguishes between the two coupling regimes is the fact
that at strong coupling, the $D2$-brane instability originates from tachyonic mode of the 
vibrational spectrum of $D2$ itself which is indeed light whereas at weak coupling, it comes
from tachyonic mode of the light $F1$'s ending on heavy $D2$. This is a new finding made in
the present work. \\
In the present work, again we remind the reader that we were particularly interested in the case
when the endpoint of both $RR$ and $NSNS$-charged brane-antibrane annihilations is the string
vacuum in which the supersymmetry is fully restored. As pointed out by Sen \cite{sen}, however,
for the case of unstable $RR$-charged $D$-branes, there is other decay channel in which the
final state is a stable BPS lower-dimensional $D$-brane with partially restored supersymmetry
instead. More precisely, as we already mentioned above, a non-BPS $D_{p}-\bar{D}_{p}$ system in 
type IIA/IIB theories may decay and settle down to a BPS $D_{(p-2)}$-brane which is a topological 
soliton emerging from the tachyon condensation via the Higgs-type mechanism \cite{yi}. 
Now that we seem to be left with two different decay channels, the first leading to the string vacuum 
with fully restored supersymmetry while the second yielding a stable BPS lower-dimensional brane
with partially restored supersymmetry. Obviously, these two decay channel are distinct since the 
endpoint of the second channel, namely a stable BPS lower-dimensional brane would never decay 
further to the vacuum spontaneously. One, then, would be puzzled and led to ask a question such as 
what would be the eventual fate of an unstable brane, or put differently, 
which channel would be favored (or more probable) over the other ? 
And the same question may be asked for the case of $NSNS$-charged brane-antibrane
decays such as the $NS5-\bar{NS5}$ system in IIA/IIB theories which can either annihilate to the
string vaccum or decay to a $D2$-brane or to a $D3$-brane respectively. The relevant answer to this
question seems to be that it would depend on which tachyon condensation mechanism happens to operate,
namely between the one leading to the string vacuum and the other yielding the topological soliton
via the Higgs-type mechanism. And for now one never knows which mechanism has better chance to
work. \\
Lastly, we stress again that either for the annihilations of the $NSNS$-charged 
brane-antibranes in IIA/IIB theories or for the decays to lower-dimensional $D$-branes 
conjectured in the literature, their physical understanding can never be complete unless the concrete 
stringy description of the quantum instability both at strong and at weak couplings is given. 
And we believe that in the present work, we have provided such a comprehensive stringy 
interpretation.

\section*{Acknowledgments}

This work was financially supported by the BK21 Project of the Korean Government. 
And the author would like to thank Sunggeun Lee for assistance to generate the figures.


\begin{thebibliography}{99}


\bibitem{non-BPS}
O.~Bergman and M.~R.~Gaberdiel, Phys.\ Lett.\ {\bf B441} (1998) 133 ;
A.~Sen, JHEP {\bf 9809} (1998) 023 ; JHEP {\bf 9810} (1998) 021 ; JHEP {\bf 9812} (1998) 021 ;
E.~Witten, JHEP {\bf 9812} (1998) 019 ;
P.~Horava, Adv.\ Theor.\ Math.\ Phys.\ {\bf 2} (1999) 1373.

\bibitem{sen}
A.~Sen, JHEP {\bf 9808} (1998) 010 ; JHEP {\bf 9808} (1998) 012 ; JHEP {\bf 9912} (1999) 027 ;
hep-th/9904207.

\bibitem{kim}
H.~Kim, Nucl.\ Phys.\ {\bf B651} (2003) 143, hep-th/0208068 ; 
        JHEP {\bf 0301} (2003) 080, hep-th/0204191.

\bibitem{lozano1}
L.~Houart and Y.~Lozano, Nucl.\ Phys.\ {\bf B575} (2000) 195 ;
JHEP {\bf 0003} (2000) 031.

\bibitem{lozano2}
K.~Intriligator, M.~Kleban and J.~Kumar, JHEP {\bf 0102} (2001) 023.

\bibitem{youm}
D.~Youm, Nucl.\ Phys.\ {\bf B573} (2000) 223.

\bibitem{bonnor}
W.~B.~Bonnor, Z.\ Phys.\ {\bf 190} (1999) 444.

\bibitem{yi}
P.~Yi, Nucl.\ Phys.\ {\bf B550} (1999) 214.

\bibitem{eric}
E.~Bergshoeff et al., Nucl.\ Phys.\ {\bf B494} (1997) 119 ; 
Class.\ Quant.\ Grav.\ {\bf 14} (1997) 2757 ;
E.~Bergshoeff, J.~Gomis and P.~K.~Townsend, Phys.\ Lett.\ {\bf B421} (1998) 109.

\bibitem{witten}
E.~Witten, Nucl.\ Phys.\ {\bf B460} (1996) 335 ; 
see also, J.~Polchinski, {\it TASI Lectures on D-branes}, hep-th/9611050.

\bibitem{ns5}
E.~Witten, hep-th/9507121 ;
A.~Strominger, Phys.\ Lett.\ {\bf B383} (1996) 44 ;
N.~Seiberg, Nucl.\ Phys.\ Proc.\ Suppl.\ {\bf 67} (1998) 158 ;
A.~Giveon and D.~Kutasov, Rev.\ Mod.\ Phys.\ {\bf 71} (1999) 983.

\end{thebibliography}
\end{document}